\documentclass[useAMS,usenatbib]{mn2e}
\usepackage{epsfig}

\title[Dust Formation and Survival in SN ejecta]{Dust Formation and Survival in Supernova Ejecta}
\author[S. Bianchi \& R. Schneider]{Simone Bianchi$^1$ and Raffaella Schneider$^2$\\
$^1$ INAF - Istituto di Radioastronomia, Sezione di Firenze, 
Largo Enrico Fermi 5, 50125 Firenze, Italy\\
$^2$ INAF - Osservatorio Astrofisico di Arcetri, 
Largo Enrico Fermi 5, 50125 Firenze, Italy}

\begin{document}

\date{3 April 2007}
\pagerange{\pageref{firstpage}--\pageref{lastpage}}
\pubyear{2007}

\maketitle
\label{firstpage}

\begin{abstract}
The presence of dust at high redshift requires efficient condensation of 
grains in SN ejecta, in accordance with current theoretical models. 
Yet, observations of the few well studied SNe and SN 
remnants imply condensation efficiencies which are about two orders of 
magnitude smaller. Motivated by this tension, we have (i) revisited the 
model of Todini \& Ferrara (2001) for dust formation in the ejecta of core 
collapse SNe and (ii) followed, for the first time, the evolution of newly 
condensed grains from the time of formation to their survival - through the 
passage of the reverse shock - in the SN remnant. We find that 0.1 - 0.6 
$M_{\odot}$ of dust form in the ejecta of 12 - 40 $M_\odot$ stellar 
progenitors. Depending on the density of the surrounding ISM, between 
2-20\% of the initial dust mass survives the passage of the reverse shock, 
on time-scales of about $4-8\times 10^4$~yr from the stellar explosion. 
Sputtering by the hot gas induces a shift of the dust size distribution 
towards smaller grains. The resulting dust extinction curve shows a good
agreement with that derived by observations of a reddened QSO at $z =6.2$.  
Stochastic heating of small grains leads to a wide distribution of dust temperatures. 
This supports the idea that large amounts ($\sim 0.1 M_\odot$) of cold dust 
($T \sim 40$K) can be present in SN remnants, without being in conflict with 
the observed IR emission. 
\end{abstract}


\section{Introduction}

In the last few years, mm and submm observations of samples of 
$5<z<6.4$ quasars have provided a powerful way of probing the very 
existence and properties of dust within 1 Gyr of the Big Bang. 
The inferred far-IR luminosities are consistent with thermal emission 
from warm dust ($T<100$~K), with dust masses $\geq 10^8 M_{\odot}$ 
\citep{BertoldiA&A2003,RobsonMNRAS2004,BeelenApJ2006,HinesApJL2006}.
Despite the uncertainties due to the poorly constrained dust temperatures 
and absorption coefficients, the estimated dust masses are huge, implying 
a high abundance of heavy elements at $z \approx 6$, consistent with the 
super-solar metallicities inferred from the optical emission-line ratios 
for many of these systems 
\citep{PentericciAJ2002,FreudlingApJL2003,MaiolinoApJL2003}. 
 
Although high redshift quasars are extreme and rare objects, hardly 
representative of the dominant star forming galaxies, the above 
observations show that early star formation leads to rapid enrichment 
of the Interstellar Medium (ISM) with metals and dust.
 
It is difficult for the dust to have originated from low-mass evolved 
stars at $z>5$ as their evolutionary timescales ($10^8$ to $10^9$ yr) 
are comparable to the age of the Universe at that time 
\citep{MorganMNRAS2003,MarchenkoProc2006}. Thus, if the observed dust 
at $z>5$ is the product of stellar processes, grain condensation in 
supernova (SN) ejecta provides the only viable explanation for its 
existence. This scenario has recently been tested through the 
observation of the reddened quasar SDSSJ1048+46 at $z=6.2$ 
\citep{MaiolinoNature2004}. The inferred extinction curve of the dust 
responsible for the reddening is different with respect to that observed 
at $z<4$ (Small Magellanic Cloud-like, \citealt{HopkinsAJ2004}), and
it shows a very good agreement with the extinction curve predicted 
for dust formed in SN ejecta.

Theoretical models, based on classical nucleation theory, predict that a few 
hundred days after the explosions silicate and carbon grains can form in 
expanding SN ejecta, with condensation efficiencies in the range 0.1-0.3
\citep{KozasaA&A1991,TodiniMNRAS2001,ClaytonApJ2001}.

Direct observational evidences for dust production have been collected only 
for a limited number of SNe, such as 1987A \citep{WoodenApJS1993}
1999em \citep{ElmhamdiMNRAS2003}, and 2003gd \citep{SugermanScience2006}.
With the exception of 2003gd, the dust masses derived from the IR emission 
are $\approx 10^{-3} M_{\odot}$, corresponding to condensation efficiencies 
which are two orders of magnitude smaller than what theory predicts. A 
fraction of dust could escape detection if it is cold and concentrated in 
clumps. This has been confirmed to be the case for SN 2003gd where a 
radiative transfer code has been used to simultaneosly fit the optical 
extinction and IR emission, leading to an estimated dust mass of 
$2 \times 10^{-2} M_{\odot}$ \citep{SugermanScience2006}.
However, when applied to SN 1987A, the same numerical model gives dust 
mass estimates which do not differ significantly from previous analytic 
results \citep{ErcolanoMNRAS2007}.

Similar low dust masses have been inferred from infrared observations of 
galactic SN remnants with Spitzer and ISO satellites 
\citep{HinesApJS2004,KrauseNature2004,GreenMNRAS2004}. The consistent 
picture that emerges is that the mid- and far-IR excess observed is due to 
emission from small amounts of warm dust, with indicative temperatures 
$T \sim 80 - 270$~K and masses $3 \times 10^{-3} - 10^{-5} M_{\odot}$  
for Cas A, and temperatures $T \sim 50$~K and masses $3 \times 10^{-3} - 
0.02 M_{\odot}$ for the Crab nebula. Cold dust has also been detected 
through far-IR and submm observations of these remnants 
\citep{DunneNature2003,KrauseNature2004}. However, the interpretation of 
these data is complicated by the strong contamination from cold dust along 
the line of sight, providing so far only upper limits of $0.2 M_{\odot}$ 
on the amount of cold dust associated to the SN remnants.
    
The aim of the present paper is to critically assess the model developed 
by \citet{TodiniMNRAS2001} exploring a wider range of initial conditions 
and model assumptions. We then follow the evolution 
of dust condensed in SN ejecta on longer timescales with respect to previous 
theoretical models. In particular, we are interested in understanding 
how the passage of the reverse shock affects the newly formed grain size 
distributions and masses, so as to make predictions for the expected dust 
properties from the time of formation in the ejecta (a few hundred days 
after the explosion) to its survival in the SN remnant, hundreds of years 
later. So far this process has received little attention, most of the 
studies being dedicated to the destruction of ISM dust grains caused by 
the SN forward shock \citep{DraineApJ1979b,JonesApJ1994,NozawaApJ2006},
with the notable exception of \citet{DwekProc2005}, who, on the basis of 
timescale considerations, finds that the reverse shock is able to destroy 
much of the initially formed dust.

The paper is organised as follows: Sect.~\ref{formation} revisits the dust
formation models based on the nucleation theory; Sect.~\ref{erosion} describes
the model adopted for the propagation of the reverse shock into the ejecta
and shows the effect of sputtering on the grain size distribution and total 
mass; in Sect.~\ref{extem} we compare the extinction and emission properties
of the surviving SN dust with observations. Finally, the results are summarised
in Sect.~\ref{summary}.

\section{SN dust formation revisited}
\label{formation}
Models of dust formation in the ejecta of core collapse SNe 
typically predict that large masses of dust ($0.1 - 1.0 M_\odot$) 
are formed within 1000 days from the onset of the explosion, when the
ejecta are still compact \citep[radius of order $10^{16}$cm;][]
{KozasaA&A1991,TodiniMNRAS2001,NozawaApJ2003}. 
If these freshly formed dust grains were distributed homogeneously 
within the ejecta, their opacity would be very high, with center-to-edge 
optical depths of order $10^2-10^4$ in optical wavelengths, depending 
on the grain material and size distribution. The ejecta would thus
be opaque to radiation produced within it \citep{KozasaA&A1991}.

Observations of recent SNe, instead, reveal extinctions smaller than 
a couple of magnitudes, which imply dust masses of only $10^{-4} - 
10^{-2} M_\odot$ \citep{SugermanScience2006,ErcolanoMNRAS2007}. The
dust mass derived from extinction measures could be underestimated if 
grains are distributed in clumps with a small volume filling factor:
for a given amount of grains, a clumpy distribution would produce a 
lower effective extinction. However, the comparison between observations
of dust extinction/emission and radiative transfer models shows that
the neglect of clumping can only produce a moderate underestimation of
the dust mass in the ejecta \citep{ErcolanoMNRAS2007}.

To check whether the dust production in SNe is overestimated, we have 
reconsidered the model of \citet{TodiniMNRAS2001}. In the model, dust 
formation is investigated in the framework of standard nucleation 
theory: when a gas becomes supersaturated, particles (monomers) aggregate 
in a seed cluster which subsequently grows by accretion of other monomers
\citep{FederAdPhy1966}. For grain materials whose molecules are not 
present in the gas phase, the {\em key species} approach is adopted 
\citep{KozasaPThPh1987}. Six materials where considered in the original 
work: amorphous carbon (AC), iron, corundum (Al$_2$O$_3$), magnetite 
(Fe$_3$O$_4$), enstatite (MgSiO$_3$) and forsterite (Mg$_2$SiO$_4$). 
Following \citet{SchneiderMNRAS2004}, we have added the formation of SiO$_2$ 
grains. SiC grains, found in meteorites and considered to be of
SN origin from their anomalous isotopic ratios \citep{ClaytonARA&A2004}, 
are not considered since their formation is impeded by the formation
of AC and Si-bearing grains \citep{NozawaApJ2003}.
The model of \citeauthor{TodiniMNRAS2001} also considers the 
formation and destruction of SiO and CO molecules: while the first is 
necessary to study the formation of Si-bearing grains, the second may 
be a sink for carbon atoms that otherwise would accrete on grains.

The ejecta are taken to have a uniform composition and density, with 
initial temperature and density chosen to match the observations of 
SN1987A.  The initial composition depends on the metallicity and mass of 
the progenitor star, $M_\mathrm{star}$, while the dynamic is given by the mass of 
the ejecta $M_\mathrm{eje}$ and the kinetic energy of the explosion 
$E_\mathrm{kin}$: the models of \citet{WoosleyApJS1995} were used. 

In the models of \citeauthor{TodiniMNRAS2001}, the gas becomes supersaturated
after a few hundred days from the explosion. The nucleation process starts
at temperature between 1800K (for AC) and 1200K (for Si-bearing materials). 
At the beginning the gas is moderately supersaturated and large seed 
clusters, made of $\cal{N}$ monomers, tend to form.  However, their 
formation rate per unit volume ({\em the nucleation current}) is small. 
As the volume of the ejecta increases, the supersaturation rate grows and 
smaller clusters aggregate with a larger formation rate. This occurs until 
the gas becomes sufficiently rarified (because of expansion and/or exhaustion 
of monomers in the gas phase) and the formation rate drops.  The nucleation 
process, together with accretion,  results in a typical log-normal grain size 
distributions \citep[see, e.g., ][]{TodiniMNRAS2001,NozawaApJ2003}.

\begin{figure}
\center{ \epsfig{file=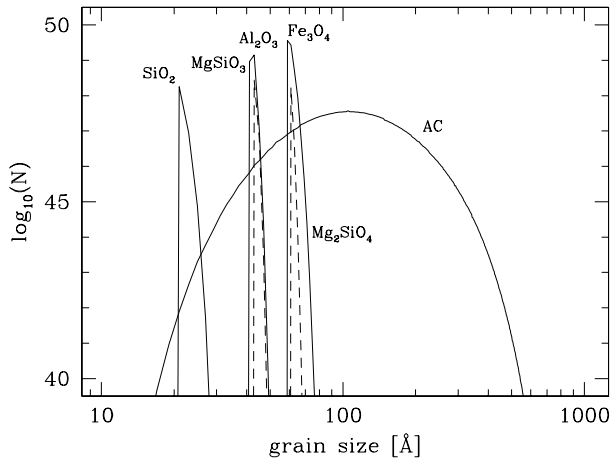,width=8cm} }
\caption{Size distribution for grains formed in the ejecta of a SN with 
a progenitor star of solar metallicity and mass $M_\mathrm{star} = 20 
M_\odot$  ($\cal{N} =\;$2). The distributions of Al$_2$O$_3$ and 
Mg$_2$SiO$_4$ are shown with dashed lines for ease of identification.}
\label{distri_s20a}
\end{figure}

For materials apart from AC, the supersaturation rate increases quickly
during the ejecta expansion, and the seed clusters can become very small. 
In \citet{TodiniMNRAS2001} seed clusters were allowed to be of any size. 
In this paper we consider only clusters with $\cal{N} \ge\;$2, 
and introduce discrete accretion of monomers. While these two (more 
physical) requirements have a limited effect on AC grains, they alter the 
size distributions and masses of grains composed by the other materials. 
In Fig.~\ref{distri_s20a} we show the size distribution of 
grains formed in the ejecta of a SN with a progenitor star of solar 
metallicity and $M_\mathrm{star}= 20 M_\odot$. Only AC grains retain the usual 
log-normal distribution. Instead, the size distribution of grains of other 
materials lacks the low-radius tail. Compared to the results of 
\citet{TodiniMNRAS2001}, their total number is reduced (since larger seed
clusters have a smaller formation rate) and their mean size is larger 
(since the monomers not allowed to form the smaller clusters are now 
available to accrete on the larger). 

It is to be noted, however, that the use of the standard nucleation theory 
is questionable when clusters are made by $\cal{N} \la\;$10 monomers 
\citep{DraineAp&SS1979,GailA&A1984}. To check what influence this limit 
has on the results, we have run models in which the formation of clusters 
with $\cal{N} <\;$10 is suppressed. The resulting size distributions 
confirm the same trend: less non-AC grains form, and of larger mean size. 
Again AC is unaffected. 

\begin{figure}
\center{ \epsfig{file=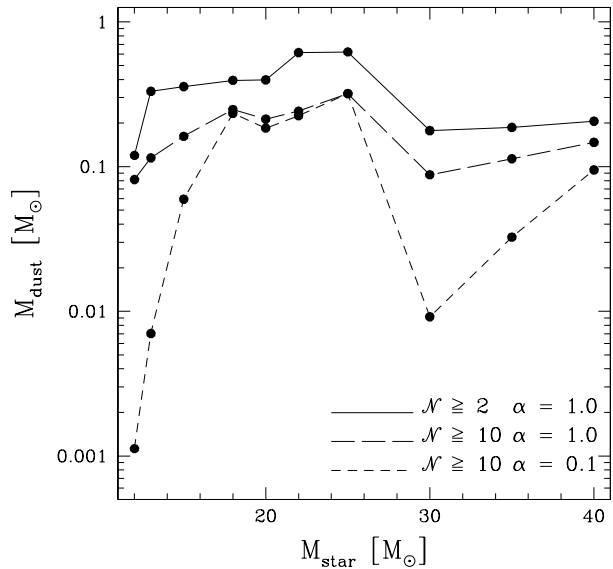,width=8cm} }
\caption{Mass of dust formed in the ejecta of a SN as a function of the 
mass of the progenitor star, for models with different minimum cluster 
size and sticking coefficient. The metallicity of the progenitors is solar.
}
\label{masses_snS}
\end{figure}

In Fig.~\ref{masses_snS} we show $M_\mathrm{dust}$, the mass of dust formed 
in the ejecta of SNe of solar metallicity, as a function of $M_\mathrm{star}$. 
The solid line refer to the models with $\cal{N} \ge\;$2. Though reduced with 
respect to \citeauthor{TodiniMNRAS2001}, still considerable masses of dust 
are formed, predominantly of AC and Fe$_3$O$_4$ grains. If $M_\mathrm{star}\le 25M_\odot$,
all the available carbon condenses in dust grains.  In the more massive 
models, roughly equal amounts of carbon goes in grains and in CO, since the 
molecule destruction mechanism provided by $^{56}$Co decay is reduced because 
of its low yield in the ejecta. Results are similar (within a factor of two) 
if the metallicity of the progenitor stars is below solar. The only 
distinction is the model with zero metallicity, where stars with 
$M_\mathrm{star}\ge 35 M_\odot$ produce no dust \citep{SchneiderMNRAS2004}. 
No substantial differences are found if a different thermal history of 
the ejecta is assumed: $M_{dust}$ is still of the same order of magnitude
if densities 
and temperatures follow the evolution adopted by \citet{NozawaApJ2003}.
As already seen in Fig.~\ref{distri_s20a}, imposing $\cal{N} \ge\;$10 
results in a great reduction of the number of non-AC grains: the dust mass 
in these models is entirely due to AC, which is unaffected by the limit 
(Fig.~\ref{masses_snS}, long-dashed line).

Dust formation models depend strongly on the sticking coefficient
$\alpha$. In most of the published models, and in the results presented 
so far, it is assumed that all gas particles colliding on a grain will 
stick to it ($\alpha = 1$). However, theory predicts that $\alpha$ 
depends on the impact energy, on the grain internal energy, and on 
the material involved: for the gas temperature at which most grains
form $\alpha$ is significantly reduced \citep{LeitchDevlinMNRAS1985}.
Indeed, laboratory experiments on the formation of cosmic dust analogs 
shows that $\alpha \approx 0.1$ for Si-bearing grains \citep{GailLNP2003}. 
Thus, we have also run models assuming  $\alpha = 0.1$ for all the species
considered.  By reducing $\alpha$, monomers
stay in the gas phase longer and dust formation is delayed to times
when super-saturation is larger: typically, smaller seed clusters form.
For $\cal{N} \ge\;$2, the number of non-AC grains is further reduced and 
their mass becomes negligible compared to that of AC. Again all available 
carbon is locked in AC grains, but their size distribution is shifted 
towards lower radii and seed clusters form with $\cal{N} <\;$10.
For $\cal{N} \ge\;$10 (Fig.~\ref{masses_snS}, dashed line) their mass 
reduces and the size distribution becomes similar to those of non-AC 
grains in Fig.~\ref{distri_s20a}. At least for low mass progenitors, 
the predicted $M_{dust}$ are closer to the values inferred by 
observations.

Clearly, the thermodynamic properties of the ejecta are at the limits of 
applicability of classical nucleation theory. A different approach may be 
needed, expecially if realistic $\alpha$ values are taken into account. 
Unless otherwise stated, in the following we will study the evolution of 
dust grains resulting from models with solar metallicity for the SN 
progenitors and $\cal{N} \ge\;$2, $\alpha=1$ (so as to conform to most 
works in literature). However, we will also discuss the results for 
models with different assuptions on $\cal{N}$ and $\alpha$.

\section{Survival in the reverse shock}
\label{erosion}

As the ejecta expands, a forward shock is driven into the ISM, which compresses 
and heats the ambient gas. The ISM becomes an hostile environment for the survival 
of dust grains preexisting the SN event, mainly because of sputtering by collisions 
with gas particles \citep{DraineApJ1979b,JonesApJ1994,NozawaApJ2006}.
In turn, the shocked ambient gas drives a reverse shock in the ejecta, 
which, by about 1000 years, has swept over a considerable fraction of its
volume. The dust within the SNe, then, has to face hostile conditions 
inside what had previously been its cradle. We study the process in this 
Section.

\subsection{Dynamics of the reverse shock}
\citet{TrueloveApJS1999} have studied the dynamics of a SN remnant through 
its nonradiative stages, the ejecta dominated and the Sedov-Taylor. They
provide analytic approximations for velocity and position of the
reverse and forward shocks, as a function of the kinetic energy 
$E_\mathrm{kin}$ and mass $M_\mathrm{eje}$ of the ejecta, and of the ISM 
density $\rho_\mathrm{ISM}$. We use here their solution for a uniform 
density distribution inside the ejecta. The values for $E_\mathrm{kin}$
and $M_\mathrm{eje}$ are the same that were used in the dust formation 
models: $E_\mathrm{kin} = 1.2 \times 10^{51}$ erg and $ 10 M_\odot \la 
M_\mathrm{eje} \la 30 M_\odot$ for stellar progenitor masses in the range 
12-40 $M_\odot$  and metallicities between zero and solar 
\citep{WoosleyApJS1995}. We study the effect of three different ISM 
environments,
with $\rho_\mathrm{ISM} = 10^{-25}$, 10$^{-24}$ and 10$^{-23}$ g cm$^{-3}$.

For each model, we have divided the ejecta into $N_\mathrm{s}$ spherical 
shells. We have assumed that all shells have the same initial width 
$\Delta R = R_\mathrm{eje}/N_\mathrm{s}$, with $R_\mathrm{eje}$ the 
initial radius of the ejecta. The mass of each shell is conserved 
throughout the evolution. For the $j$ shell (counting shells outwards),
the initial velocity of the gas at its inner boundary is given by 
homologous expansion,
\begin{equation}
v_j=v_\mathrm{eje}\frac{R_j}{R_\mathrm{eje}},
\qquad v_\mathrm{eje}=\sqrt{\frac{10}{3} \frac{E_\mathrm{kin}}{M_\mathrm{eje}}},
\label{eq_homo}
\end{equation}
where $R_j$ is the initial radius of the inner shell boundary and
$v_\mathrm{eje}$ is the velocity of the external boundary for ejecta
of uniform density. For practical
purposes, we start our simulation at a time $t_0$ (ideally, $t_0 
\rightarrow 0$), and we set $R_\mathrm{eje}=v_\mathrm{eje} t_0$. The 
results do not depend on the exact value of $t_0$, provided it is taken 
small enough (we use a value of order a few tens of years).

After setting the initial conditions, we study the evolution of the 
ejecta with time. At each time step, the reverse shock goes inward 
through a single shell. Thus, at time $t_i$, the reverse shock has
travelled inward through $i$ shells, and lies at the inner boundary 
of shell $j_\mathrm{rs}=N_\mathrm{s}-i-1$. Shells that have not been 
visited by the reverse shock (for $0\le j \le j_\mathrm{rs}-1$) 
continue to follow homologous expansion, i.e. the inner and outer radii 
grow linearly with time, with velocity given by Eq.~\ref{eq_homo}. 
Following the shell expansion (increase in the shell volume $V_j$), the 
shell gas density decreases as $\rho_j \propto V_j^{-1}$. For an adiabatic 
expansion, the shell temperature scales as $T_j \propto V_j^{1-\gamma}$,
with $\gamma=5/3$ \citep{TrueloveApJS1999}. Since the shock is strong,
the results are independent on the initial choice for the gas 
temperature in the ejecta. 

For the shell $j=j_\mathrm{rs}$ that has been swept over by the shock 
at time $t_i$, we apply the standard Rankine-Hugoniot jump conditions for 
a strong adiabatic shock. The density, velocity and temperature change as
\[
\rho_j = \frac{\gamma+1}{\gamma-1} \rho'_j,
\]
\[
v_j = v'_j-\frac{2}{\gamma+1} \tilde{v}_\mathrm{rs},
\]
\[
T_j = 2 \frac{\gamma-1}{(\gamma+1)^2}\frac{m}{k} \tilde{v}_\mathrm{rs}^2,
\]
where $\rho'_j$ and $v'_j$ are the density and velocity before the shock 
(i.e. following the same evolution as for shells with $j<j_\mathrm{rs}$),
$\tilde{v}_\mathrm{rs}$ is the velocity of the reverse shock in the reference 
frame of the unshocked ejecta (provided by \citealt{TrueloveApJS1999}),
$m$ is the mean particle mass and $k$ the Boltzmann's constant. To ensure
mass conservation, the volume of shell $j=j_\mathrm{rs}$ is reduced by a
factor $(\gamma-1)/(\gamma+1)$.

\begin{figure}
\center{ \epsfig{file=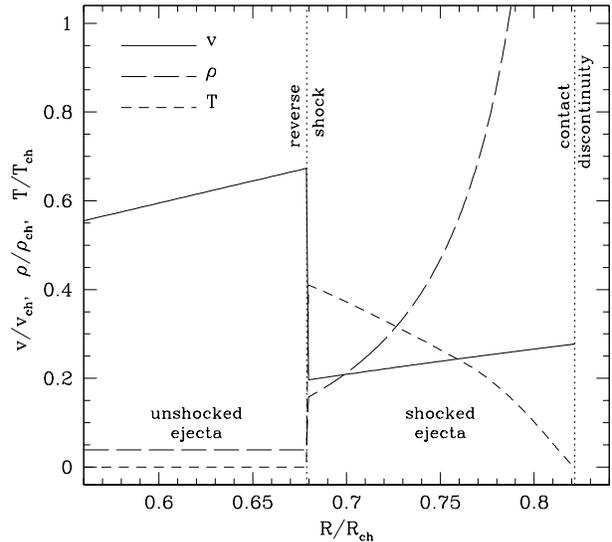,width=8cm} }
\caption{Velocity, density and temperature of the ejecta at 
$t=t_\mathrm{ch}$ as a function of the radius, for the model with 
progenitor star of solar metallicity and mass $M_\mathrm{star} = 
20 M_\odot$ ($M_\mathrm{eje}=18 M_\odot$) expanding in a ISM with 
density $\rho_\mathrm{ISM}=10^{-24}$ g cm$^{-3}$.  All quantities are 
normalised to their {\em characteristic} values \citep{TrueloveApJS1999}. 
For the model shown here $t_\mathrm{ch}=5800$~yr, $R_\mathrm{ch}=10.7$~pc, 
$T_\mathrm{ch}= 5.7\times 10^7$~K, $v_\mathrm{ch}=1800$~km s$^{-1}$. It is
also $\rho_\mathrm{ch}=\rho_\mathrm{ISM}$. The {\em contact discontinuity}
marks the border between the shocked ejecta and the ISM swept by the
forward shock.
}
\label{dynamics}
\end{figure}

For the shells $j_\mathrm{rs} < j < N_\mathrm{rs}$ shocked at earlier times
$t < t_i$, we compute the velocity $v_j$ by interpolating between the
velocity of the $j=j_\mathrm{rs}$ shell and the velocity of the forward shock 
(in the ambient rest frame), as a function of the logarithm of the shell
inner radius. Velocity and position of the forward shock are also given
by \citet{TrueloveApJS1999}. As for shells with $j<j_\mathrm{rs}$, the 
evolution of density and temperature is derived from the condition of 
adiabatic expansion and conservation of the shell mass. The typical trends
for velocity, density and temperature around the reverse shock are shown
in Fig.~\ref{dynamics}.

We checked the results (in particular the assumption for the
evolution of $j>j_\mathrm{rs}$ shells) with the 1-D hydrodynamical models
of SN blast waves of \citet{VanderSwaluwA&A2001} and with simulations kindly 
provided by L. Del Zanna (based on the code described in 
\citealt{DelZannaA&A2003}). However crude, our approximation provide a 
simple and fast solution for the density and temperature evolution of 
the ejecta during the passage of the reverse shock. Chosing an adequate
number of shells (we use $N_\mathrm{s} = 400$), it agrees with the
complete hydrodynamical solution within a factor of 2. 

\subsection{Dust grain survival}
We assume that dust grains are distributed uniformly within the ejecta, 
and that the size distribution is the same everywhere.
In the shells that have been visited by the reverse shock, dust grains
are bathed in a gas heated to high temperature (of order $10^7 - 10^8$ K
for the cases studied here). Also, the gas is slowed down and dust grains 
decouple from it, attaining a velocity relative to the gas
\[
v_{\mathrm{d}j}= \frac{2}{\gamma+1} \tilde{v}_\mathrm{rs}.
\]
Gas particles thus impact on dust grains transferring thermal and
kinetic energy, which are of the same order of magnitude
(both depending on the reverse 
shock velocity $\tilde{v}_\mathrm{rs}$, which is of order 10$^3$ km 
s$^{-1}$). Thermal and non-thermal sputtering result, which erode the 
dust grain, reducing its size. Eventually, the gas drag due to direct 
and Coulomb collisions slows the grain and non-thermal sputtering weakens. 
In this work we consider both thermal 
and non-thermal sputtering, but we neglect the gas drag and the grain charge: 
once passed through the reverse shock, the grain retains its velocity relative 
to the gas. We can thus provide upper limits on the influence of non-thermal sputtering.

The number of atoms that are sputtered off a dust grain per unit time is 
given by the sputtering rate $dN/dt$, a complex function of the gas density, 
temperature and of the nature of the dust/gas (target/projectiles) 
interaction (full expressions for $dN/dt$ can be found elsewere, see e.g.\ 
\citealt{BianchiMNRAS2005}). The sputtering rate depends on the sputtering 
yield, $Y$, the fraction of atoms that leave the target per projectile 
collision, which is a function of the energy of the impact. We use here the $Y$ 
functions described in \citet{NozawaApJ2006}, and we consider collisions 
of dust grains with H, He and O atoms in the ejecta. The grain radius
decreases with sputtering as
\begin{equation}
\frac{da}{dt} = - \frac{a_\mathrm{m}^3}{3 q a^2} \frac{dN}{dt},
\label{eq_dadt}
\end{equation}
where, $q$ is the number of atoms in a molecule of the grain material, and 
$a_\mathrm{m}$ is the {\em molecule radius}, computed from the material 
density and the molecule mass. The values for $a_\mathrm{m}$ can be derived 
easily from the $a_{0j}$ values of Table~2 in \citet{NozawaApJ2003}.
At each time step, we reduce the grain size according to Eq.~\ref{eq_dadt}
in all shells that have been swept by the reverse shock. We follow the
evolution until the reverse shock arrives near the center of the ejecta:
this is the limit of validity of the approximations in 
\citet{TrueloveApJS1999}. After that, we simply assume that the ejecta
expands adiabatically, and we end the simulations when the sputtering rate 
becomes negligible.
Since we do not include gas drag and grain charge, grains do not attain
differential velocities for different sizes. Thus, we have neglected
destruction due to grain-grain collisions. However, sputtering dominates
over this process for the high shock velocities considered here 
\citep{JonesApJ1994}.

\begin{figure}
\center{ \epsfig{file=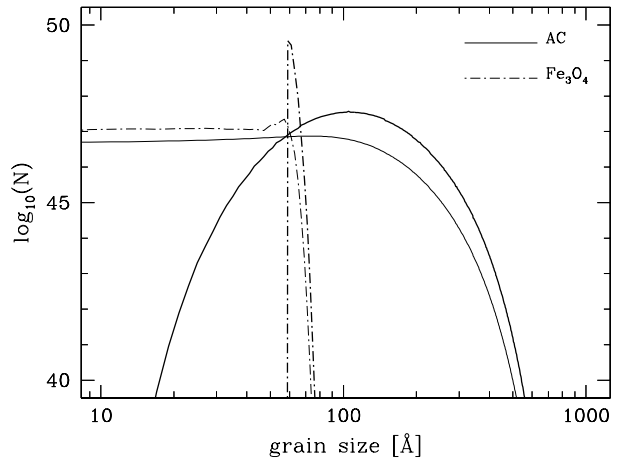,width=8cm} }
\caption{Changes in the size distribution of AC and Fe$_3$O$_4$ grains.
For each material, the thick line is the initial size distribution (the
same as in Fig.~\ref{distri_s20a}).  The thin line is the size distribution 
after the passage of the reverse shock through the ejecta.
}
\label{distriafter}
\end{figure}

Dust grains in the ionized shocked gas are heated mainly by collisions with 
electrons. If the grains are small, heating is stochastic and an 
equilibrium temperature does not exist. Instead, a broad temperature distribution 
$P(T_\mathrm{d})$ establishes, peaking at low temperature but extending also to 
high values \citep{DwekApJ1986}. The temperature may be so high that dust grains 
sublimate \citep{GuhathakurtaApJ1989}. For the cases studied here, however, 
sublimation is negligible. Details on the calculation are presented in 
Appendix~\ref{stocha}.

In Fig.~\ref{distriafter} we show the initial (thick lines) and final 
(thin lines) size distributions for AC and Fe$_3$O$_4$ grains in
the ejecta of a star with $M_\mathrm{star}= 20 M_\odot$ expanding in a medium 
with $\rho_\mathrm{ISM}=10^{-24}$ g cm$^{-3}$. As it is evident for 
the (initially) more peaked size distributions of magnetite, sputtering 
produces a {\em leaking} towards smaller sizes. The evolution of
the size distribution is analogous to that of ISM grains destroyed 
by the forward shock \citep{NozawaApJ2006}.

\begin{figure}
\center{ \epsfig{file=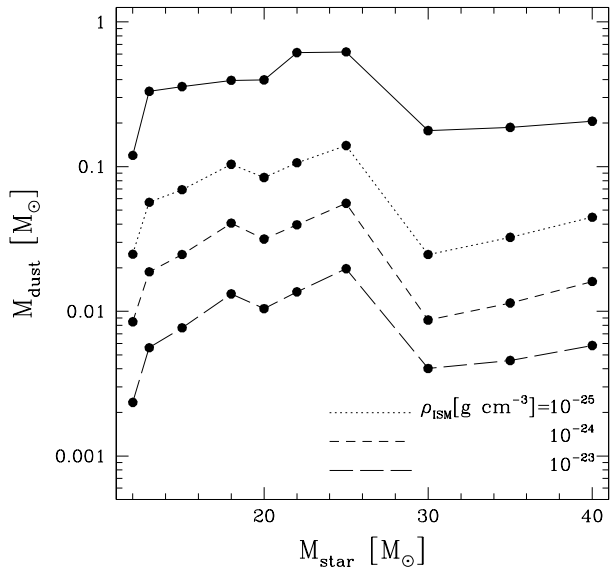,width=8cm} }
\caption{Mass of dust that survives the passage of the reverse shock in the 
ejecta, as a function of the mass of the progenitor star and of the density 
of the surrounding ISM. The solid line shows the initial dust mass
(same as the solid line in Fig.~\ref{masses_snS}). }
\label{masses_ism}
\end{figure}

In Fig.~\ref{masses_ism} we show the mass of dust that survives the passage 
of the reverse shock. For the reference model (dashed line), the erosion 
caused by sputtering reduces the dust mass to about 7\% of its initial value, 
almost independently of the stellar progenitor model. 
Most of the dust (about 70\% in mass) is consumed within one 
{\em characteristic time} $t_\mathrm{ch}$ from the explosion, when 95\% 
of the original volume of the ejecta has been reached by the reverse 
shock \citep[$t_\mathrm{ch} = 4-8 \times 10^{4}$ yr for the ejecta discussed 
here;][]{TrueloveApJS1999}. Dust in the inner shells is less affected
by erosion, because the sputtering rate is lower. A minor fraction of the 
dust mass, less than 10\%, is consumed after the reverse shock bounces at the 
center of the ejecta \citep[for $t\ga 2.6 t_\mathrm{ch}$;][]{TrueloveApJS1999}.

If the SN explodes in a denser ISM, the reverse shock would travel faster
inside the ejecta and would encounter a gas at higher density. This increases
the effect of sputtering. In Fig.~\ref{masses_ism} (long-dashed line) we see 
the fraction of dust mass that survives when $\rho_\mathrm{ISM}= 10^{-23}$ g 
cm$^{-3}$: only about 2\% of the dust mass survives. Conversely, for a lower 
density ISM, a larger fraction is left: for $\rho_\mathrm{ISM}=10^{-25}$ g 
cm$^{-3}$, it is 20\% (dotted line). While the number of surviving grains
changes with the ISM density, the shape of the size distributions remain
similar in all cases, with the typical patterns shown in Fig.~\ref{distriafter}.

No substancial change is observed in models where the dust was produced by
progenitors of metallicity different from solar. Dust destruction is instead 
enhanced in models where a smaller sticking coefficient is adopted. If
$\alpha = 0.1$ (Sect.~\ref{formation}), only 10, 3 and 1\% of the original 
dust mass survives, respectively, for $\rho_\mathrm{ISM} = 10^{-25}$, 
10$^{-24}$ and 10$^{-23}$ g cm$^{-3}$ (compared to 20, 7 and 2\% for
$\alpha = 1.0$). This is because for smaller values of $\alpha$, the dust 
distribution is made by grains of smaller radii, which are more easily
destroyed.

\section{Extinction and emission from SN dust}
\label{extem}

\citet{MaiolinoNature2004} measured the reddening in the rest-frame UV spectrum 
of a $z=6.2$ QSO and found it to be different from that of the SMC, typically used 
to deredden the spectra of lower redshift QSOs. The measured reddening is instead
compatible with the extinction law from the
\citet{TodiniMNRAS2001} SN dust model. We repeat here the same analysis using 
the updated dust formation models of Sect.~\ref{formation} and the final 
distributions after the reverse shock passage of Sect.~\ref{erosion}. As in
\citet{MaiolinoNature2004}, we derive the extinction properties from the grain 
sizes using the \citeauthor{MieAnnPhys1908}'s (\citeyear{MieAnnPhys1908}) 
theory for spherical dust grains 
and refractive indexes for dust materials from the literature 
(references are provided in Table~\ref{tab_stocha}). The procedure is analogous 
to that adopted by \citet{HirashitaMNRAS2005} when modelling the dust extinction 
from the SNe dust models of \citet{NozawaApJ2003}.

\begin{figure}
\center{ \epsfig{file=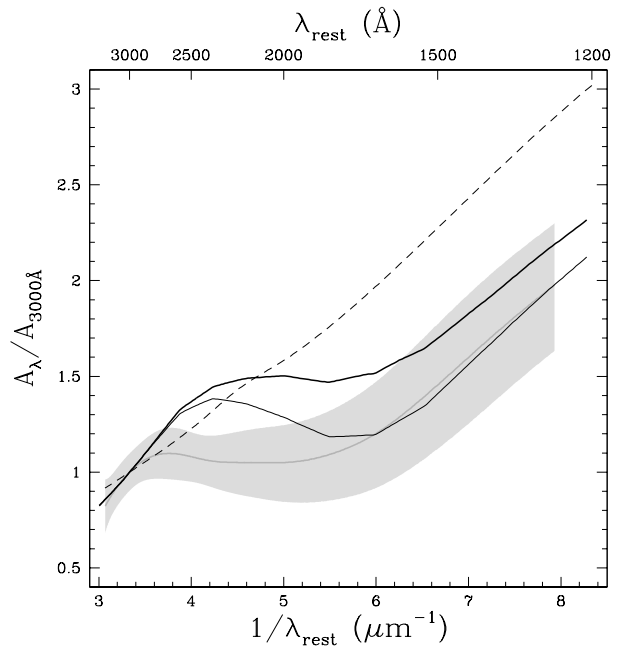,width=8cm} }
\caption{Extinction law for SN dust. The thick solid line is the extinction 
law for dust freshly formed in the ejecta. The thin solid line is the extinction 
law from dust processed by the reverse shock. The curves are computed from the 
IMF-averaged size distributions of grains formed in SNe from solar metallicity 
progenitors (see text for details). The gray line and shaded area are the 
extinction law measured on a $z=6.2$ QSO and its uncertainty 
\citep{MaiolinoNature2004}.  The dashed line is the extinction law of the 
SMC \citep{PeiApJ1992}.
}
\label{extlaw}
\end{figure}

In Fig.~\ref{extlaw} we show the results for dust formed in SNe from progenitors 
of solar metallicity. The grain size distributions from progenitors of different 
masses have been averaged over a stellar Initial Mass Function (IMF): we adopted
the Salpeter IMF, but the results do not depend heavily on this 
choice \citep{MaiolinoNature2004}. The thick solid line represents the extinction
law of dust as formed within the ejecta, without taking into account the grain
processing caused by the reverse shock. The SN dust extinction law is still 
flatter than the SMC extinction law, but the agreement with the observations 
(shaded area) is worse than in \citet{MaiolinoNature2004}. This is mainly due
to a change in the grain materials that contribute to extinction: apart from
AC, present in both the old and new model, the rise at $\lambda < 2000$ \AA\ was 
due to Mg$_2$SiO$_4$ grains, with a minor contribution from Fe$_3$O$_4$. 
In the new model, Mg$_2$SiO$_4$ contribution is insignificant, while Fe$_3$O$_4$
grains (larger than in the original model) cause the far UV rise. The bump at
$\lambda \approx 2500$ \AA\ is due to AC grains and it is typical of the optical 
properties derived from amorphous carbon formed in an inert athmosphere (the 
{\em ACAR} sample of \citet{ZubkoMNRAS1996}).

During the passage of the reverse shock, Fe$_3$O$_4$ grains are consumed more
effectively than AC grains. The resulting extinction law (thin solid line in 
Fig.~\ref{extlaw}) becomes flatter, leading to an excellent agreement with 
observations at $\lambda \le 1600$ \AA.
These results apply for ejecta expanding in a medium with  
$\rho_\mathrm{ISM} = 10^{-24}$ g cm$^{-3}$. There is no significant change in 
the extinction law if different ISM densities are considered, since the size 
distributions are similar in all cases (though the extinction at any given
wavelength is smaller for higher $\rho_\mathrm{ISM}$, because less grains 
survive). It is worth noting that grains with $a \la 20$ \AA\, though as 
abundant as larger
grains, do not contribute to the extinction law because of their reduced 
extinction cross section. As in \citet{MaiolinoNature2004}, we find that 
if progenitors of metallicity lower than solar are considered, the difference 
in the resulting extinction laws are small and lie within 0.1 y-axis units 
from the lines plotted in Fig.~\ref{extlaw}.

Extending calculations to the infrared, we have derived the dust emissivity. 
For all the IMF averaged size distributions, the emissivity in the wavelength
range $10 \le \lambda/\mu m \le 1000$ is rather featureless, and can be well
described by a power law in wavelength of index -1.4 with 
$\kappa(100\mu m) = 40$ cm$^2$ g$^{-1}$ for models where all dust has been
processed by the reverse shock. Emissivities for dust formed from progenitors 
of a given mass are within 10\% of the IMF averaged value, while the emissivity
at the end of dust condensation, before any significant sputtering has occurred,
is found to be about 20\% lower. No significant dependence is found on the metallicity
of the progenitor and on $\rho_\mathrm{ISM}$. In all cases, the emissivity is
almost entirely due to the large AC grains\footnote{For the same reason, 
increasing the minimum cluster size $\cal{N}$ and/or decreasing the sticking 
coefficient $\alpha$ does not affect the predicted extinction laws and 
emissivities, which are similar to those found for our reference model 
after the passage of the reverse shock.}.

The amount of shock-heated dust in the ejecta can be derived from infrared 
observations of SN remnants. A notable (and debated) case is that of Cas A,
the remnant from an historical SN which shows infrared emission from the 
region between the forward and reverse shocks. The identity of Cas A's 
progenitor is still highly debated. A star of 15-25 $M_\odot$ that loses its
hydrogen envelope through winds \citep{ChevalierProc2006} or binary interactions
\citep{YoungApJ2006} and then undergoes an energetic explosion can match all the
available observational constraints. In particular, the age and dynamics 
suggest a mass for the ejecta of 3 $M_\odot$, with about the same amount of gas 
reached by the reverse shock in the ejecta and swept by the forward shock in the 
surrounding ISM \citep{TrueloveApJS1999}. Given these uncertainties, and the 
dependence of the predicted dust masses on the stellar progenitor 
(see Fig.~\ref{masses_ism}), we can only give a tentative estimate of the
amount of dust predicted for Cas A by our model.     
An ejecta evolution compatible with observations can be obtained for 
a 12 $M_\odot$ progenitor, provided we neglect the hydrogen mass. In such a 
model, $\approx$0.1 $M_\odot$ of dust forms. By the age of the remnant 
($\sim$325 yr), $\approx$0.05 $M_\odot$ survives in the region reached by 
the reverse shock, where it is heated by the hot gas. 
We also need to consider the contribution to emission from dust in the ISM 
reached by the forward shock. Typically, dust in the shocked ISM is exposed 
to a gas of similar density and temperature to those in the reverse shock 
\citep{VanderSwaluwA&A2001}. For a standard value of the ISM gas-to-dust 
mass ratio, one would roughly expect a similar mass of emitting dust in 
the ISM. Thus, a model for Cas A remnant would have about 0.1 $M_\odot$
of emitting dust.

\begin{figure}
\center{ \epsfig{file=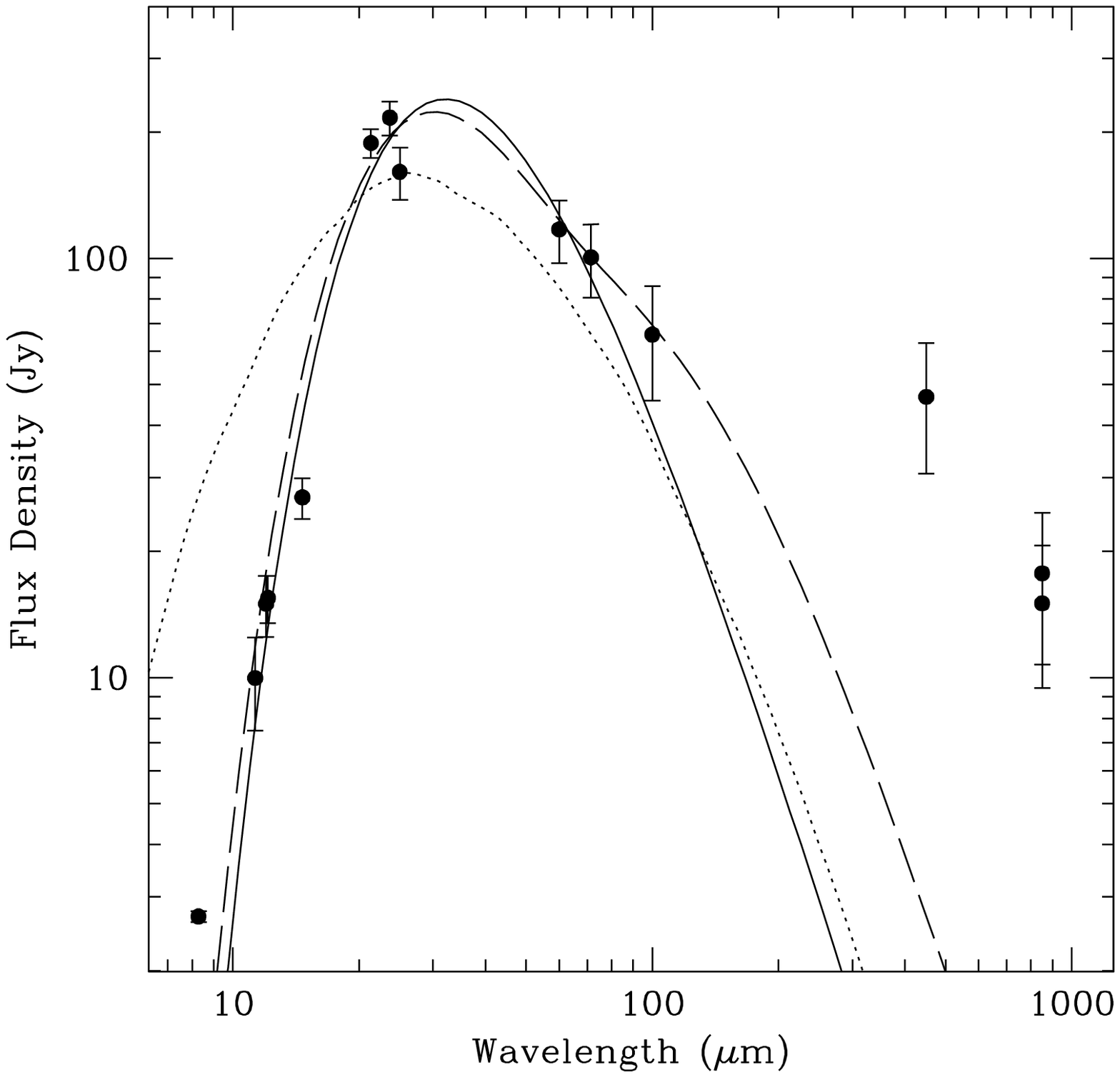,width=8cm} }
\caption{Synchrotron-subtracted SED of dust emission in CasA. Data points 
are from \citet{HinesApJS2004}. The solid line is a one-component modified 
blackbody fit to the data for $\lambda \le 100 \mu$m ($T\approx 100$K, 
$M_\mathrm{d}\approx 4.0 \times 10^{-3} M_\odot$). The dashed line is a 
two-component fit with $T\approx 110$K, $M_\mathrm{d}\approx 3.0 \times 
10^{-3} M_\odot$ and $T\approx 35$K, $M_\mathrm{d}\approx 0.1 M_\odot$. 
The dotted line is the spectrum from stochastically heated dust in our 
model. See text for details.}
\label{CasA}
\end{figure}

This mass appears to be more than an order of magnitude larger than what could be 
derived fitting the observed Spectral Energy distribution (SED) of Cas A 
\citep{HinesApJS2004}. Using the emissivity
predicted for SN dust, the flux in the wavelength range $10 \le \lambda/\mu m \le 100$ 
can be reasonably well reproduced with a single modified blackbody with temperature 
$T=100$K, and a dust mass of $4\times 10^{-3}$ $M_\odot$ (Fig.~\ref{CasA}, solid line). 
\citet{HinesApJS2004} obtain for the cold, more massive component a similar dust mass 
with $T=80$K. However, the large uncertainties and the limited FIR coverage allow to fit, 
equally well, a two-component model with temperatures 110 and 35K and masses, respectively, 
of $3\times 10^{-3}$ and 0.1 $M_\odot$ (Fig.~\ref{CasA}, dashed line). 
Unfortunately Cas A lies on the line of sight of dense molecular clouds which do not allow 
a reliable estimate of the cold dust mass from observations at longer wavelengths in the 
FIR and sub-mm. Still, upper limits on the dust mass in the remnant are compatible with 
our model predictions \citep{KrauseNature2004}. 

A broad span of temperatures is clearly needed for a reliable estimate 
of the dust mass in the remnant. In Fig.~\ref{CasA} we also show the 
SED of the shock heated dust in the CasA model (dotted line). Because 
of stochastic heating (Appendix~\ref{stocha}), grains have temperatures
mainly ranging from 10 to 100K. The SED cannot be easily modelled
using a 2-component modified blackbody: the longer wavelength side 
could be described with a cold component of $T\approx 60$K, which would
underestimate the dust mass by about a factor 5; instead, a hot 
component at $T\approx 150$K would leave a substantial residual
in the fit at $\lambda \la 10 \mu$m. When comparing to the data
for CasA, it appears that dust temperature in our models is 
overestimated. This could be due to an overestimate of the dust
stochastic heating, to a reduction of smaller grains with respect
to the dust formation model, or to differences between the 
emission properties of true and modelled materials.
However, the uncertainties in the thermal/dynamical history of 
the ejecta of CasA and the impossibility of discriminating
between ISM and ejecta dust emission in the spectrum prevent
a more detailed analysis.

\section{Summary}
\label{summary}

In the present work we have revisited the model of Todini \& Ferrara (2001)
for dust formation in the ejecta of core collapse SNe and followed the 
evolution of newly condensed grains from the time of formation to their survival 
through the passage of the reverse shock. 
 
The main results can be summarized as follows:
\begin{enumerate}
\item The new features introduced in the dust formation model have only a minor impact on 
AC grains but significantly affect other species (Si-bearing grains, Al$_2$O$_3$, and Fe$_3$O$_4$).
For 12 - 40 $M_\odot$ stellar progenitors with $Z=Z_\odot$, the predicted M$_\mathrm{dust}$ ranges between 
0.1 - 0.6 $M_{\odot}$; comparable values (within a factor 2) are found if the progenitors 
have $Z<Z_\odot$. The dominant grain species are AC and Fe$_2$O$_3$. 

\item We identify the most critical parameters to be the minimum number of monomers, ${\cal N}$, which define a critical seed 
cluster, and the value of the sticking coefficient, $\alpha$. Assuming ${\cal N} \ge 10$ (below which the application of standard
nucleation theory is questionable) results in a great reduction of non-AC grains because these species
nucleate when the gas in the ejecta is highly super-saturated and smaller                           
seed clusters form. This effect is further enhanced if $\alpha < 1$: for $\alpha = 0.1$ and 
stellar progenitor masses $M_\mathrm{star} < 20 M_{\odot}$, the total mass of dust is reduced 
to values in the range 0.001-0.1 $M_{\odot}$, comparable to those inferred from the IR emission at 
400-700 days after the explosion for 1987A and 2003gd, the only two core-collapse SNe for which 
these data were available. 

\item Using a semi-analytical model to describe the dynamics of the reverse shock, we have found 
that thermal and non-thermal sputtering produce a shift of the size distribution function 
towards smaller grains; the resulting dust mass reduction depends on the density of the 
surrounding ISM: for $\rho_\mathrm{ISM} = 10^{-25}, 10^{-24}, 10^{-23}$ g cm$^{-3}$, 
about 20\%, 7\%, and 2\% (respectively) of the initial dust mass survives. Most of dust 
consumption occurs within one characteristic time from the explosion, about $4-8\times 10^4$~yr 
for core-collapse SNe. Thus, the impact of the reverse shock needs to be taken into account 
when comparing model predictions with observations of young SN remnants. 
  
\item Averaging over a Salpeter IMF, we have derived dust extinction and emissivity. 
We find that the extinction curve is dominated by AC and Fe$_3$O$_4$ grains with radii larger than 20 \AA . 
As a result, it is relatively flat in the range 1500-2500\AA\, and then rises in the far UV.
Thus, the peculiar behaviour of the extinction produced by SN dust, which has been successfully used to 
interpret observations of a reddened QSO at $z=6.2$ \citep{MaiolinoNature2004}, is preserved in the 
present model, and it is further amplified by the modifications induced by the passage of the 
reverse shock. 

\item Using dust emissivity predicted by the model, we can reproduce the observed IR 
flux from the young SN remnant CasA adopting a single modified black-body of temperature 
$T=100$~K, which implies a mass of warm dust of $4\times 10^{-3} M_\odot$, consistent with \citet{HinesApJS2004}.     
However, the limited observational coverage in the FIR allows to equally well reproduce the data
adding a cold component with temperature $T=35$~K and dust mass of 
$0.1 M_\odot$. According to our model, such a mass of dust is what would be produced by a
single 12 $M_\odot$ star that has exploded after losing its hydrogen envelope, a plausible 
candidate for the highly debated CasA's progenitor. Because of the stochastic heating of small 
grains by collisions with hot gas electrons, dust in the shocked gas is predicted to have 
temperatures ranging from $10$ to $100$K. 
\end{enumerate}

We conclude that our study supports the idea that core-collapse SNe can be major dust factories. 
At the same time, it shows that our knowledge of dust condensation and its survival in SN ejecta
still lacks to control some critical parameters, which prevent reliable estimates of condensation 
efficiencies, especially for the less massive progenitors. Within these uncertainties, the model
can accomodate the still sparse observational probes of the presence of dust in SN and SN remnants.

\section*{Acknowledgments}
We are grateful to A. Ferrara for profitable discussions and suggestions, 
and to L. Del Zanna for kindly providing us the results of 1-D hydrodynamical 
simulations.  We also acknowledge DAVID 
members\footnote{http://www.arcetri.astro.it/science/cosmology} for fruitful 
comments and Cristiano Porciani for precious help.

\bibliography{/home/voltumna/sbianchi/tex/DUST}

\appendix

\section{Stochastic heating from electron collisions} 
\label{stocha}

We have derived the temperature distribution $P(T_\mathrm{d})$ following 
the method of \citet{GuhathakurtaApJ1989}, to which we refer for a more 
detailed description.

We have divided the range of possible dust temperatures into $N_\mathrm{b}$
bins. For the case studied here, we found sufficient to define 
$N_\mathrm{b} = 500$ bins, logarithmically spaced in the range 
$2 < T_\mathrm{d} / $K$ < 2000 $. The $i$-bin has temperature 
$T_{\mathrm{d},i}$, energy $E_i$ and energy width $\Delta E_i$.
The energy corresponding to each
value of $T_\mathrm{d}$ is defined as
\[
E(T_\mathrm{d}) = \int_0^{T_\mathrm{d}} C(T) dT,
\]
where $C$ is the specific heat, which is derived by fitting experimental data.
We have adopted a piecewise power-law of the form
\[
C(T)= A\; T^B.
\]
The fitted values for $A$ and $B$ are given in Table~\ref{tab_stocha}. 

\begin{table*}
%\begin{minipage}{20cm}
\caption{Thermal and optical properties of SNe dust materials}
\label{tab_stocha}
\begin{tabular}{lcccll}
Materials & \multicolumn{4}{c}{Specific heat} & 
\multicolumn{1}{c}{Refractive index}\\
          & A (erg cm$^3$ K$^{-1}$) & B & range in T (K) & 
\multicolumn{1}{c}{Refs.} & \multicolumn{1}{c}{Refs.} 
\\ \hline
Al$_2$O$_3$      & 4.44        & 3     & $(0,110]  $ &
\citet{DitmarsJRNBS1982}& \citet[{\em ISAS}]{KoikeIcarus1995}\\
                 & 1.22 10$^2$ & 2.29  & $(110,200]$ &
\citet{ChaseJANAF1985}& \citet[{\em compact}]{BegemannApJ1997}\\
                 & 7.00 10$^5$ & 0.66  & $(200,500]$ &\\
                 & 1.45 10$^7$ & 0.17  & $(500,2000]$ &\\
                 & 5.44 10$^7$ & 0     & $(2000,\infty)$ &\\
\\
Fe$_3$O$_4$      & 22.5        & 3     & $(0,10]$  &
\citet{ChaseJANAF1985}& \citet{MukaiProc1989}\\
                 & 7.65        & 3.47  & $(10,30]$  &
\citet{ShepherdPhysRevB1985}& \\
                 & 3.74 10$^2$ & 2.32  & $(30,80]$  &
\citet{KoenitzerPhysRevB1989}& \\
                 & 1.04 10$^5$ & 1.04  & $(80,300] $  &
                 & \\
                 & 3.95 10$^7$ & 0     & $(300,\infty)$ &\\
\\
MgSiO$_3$        & 9.22        & 3     & $(0,20]$  & 
\citet{KelleyJACS1943}& \protect{\citet{JaegerA&A2003}}\\
                 & 2.01        & 3.51  & $(20,50]$  & 
\citet{ChaseJANAF1985}& \\
                 & 7.94 10$^2$ & 1.98  & $(50,130]$  & 
\citet{KrupkaAmMin1985}& \\
                 & 1.32 10$^5$ & 0.93  & $(130,400]$  & \\
                 & 3.47 10$^7$ & 0     & $(400,\infty)$ &\\
\\
Mg$_2$SiO$_4$    & 22.7        & 3     & $(0,60]$ & 
\citet{KelleyJACS1943}& \protect{\citet{JaegerA&A2003}}\\
                 & 6.45 10$^2$ & 2.18  & $(60,120]$  &
\citet{ChaseJANAF1985}& \\
                 & 1.40 10$^5$ & 1.06  & $(120,300]$  &\\
                 & 1.42 10$^7$ & 0.25  & $(300,2000]$  &\\
                 & 9.44 10$^7$ & 0     & $(2000,\infty)$ &\\
\\
AC               & 3.82 10$^2$ & 2     & $(0,70]$  &
\citet{ChaseJANAF1985}& \citet[{\em ACAR}]{ZubkoMNRAS1996}\\
                 & 3.27 10$^3$ & 1.50  & $(70,300]$  &
\citet{DraineApJ2001}& \\
                 & 5.58 10$^4$ & 1.00  & $(300,700]$  &\\
                 & 1.09 10$^7$ & 0.19  & $(700,3000]$  &\\
                 & 5.09 10$^7$ & 0     & $(3000,\infty)$ &\\
\\
SiO$_2$          & 9.95 10$^2$ & 2     & $(0,60]$  &
\citet{ChaseJANAF1985}& \citet{PhilippHOCS1985} \\
                 & 5.50 10$^4$ & 1.02  & $(60,500]$  & 
\protect{\citet{LegerA&A1985}}& 
\protect{\citet{HenningA&A1997}}\\
                 & 3.11 10$^7$ & 0     & $(500,\infty)$ &\\
\hline
\end{tabular}
%\end{minipage}
\end{table*}

The function $P(T_\mathrm{d})$ is computed from the transition matrix
between the initial and final internal energy states, $A_{f,i}$. The 
probability per unit time that a dust grain in the energy state $E_i$ 
is heated to the energy state $E_f$ (with $f > i$) by collisions with
electrons is given by
\[
A_{f,i} = \left\{
\begin{array}{l}
n\;\pi a^2 \Delta E_f \times \\
\\
\Bigg[\sqrt{\displaystyle \frac{2 (E_f-E_i)}{m}} f(E_f-E_i)+
\quad\mbox{if}\;\; E_f-E_i < E_\star\\
\\
\sqrt{\displaystyle \frac{2 E_\mathrm{eff}}{m}} f(E_\mathrm{eff})
\frac{\displaystyle \sqrt{1-{(E_f-E_i)}/{E_\mathrm{eff}}}}
{\displaystyle 1-   \sqrt{1-{(E_f-E_i)}/{E_\mathrm{eff}}}
} \Bigg], \\
 \\
 \\
0 \qquad \qquad\mbox{otherwise}.
\end{array}
\right.
\]
In the above equation, $n$ and $m$ are the electron number density and 
mass, respectively, and  $f(E)$ is the Maxwell distribution function defined
such that $n f(E) \Delta E$ is the number of electrons of energy $E$
per unit volume.
$E_\star$ is the maximum energy that an electron can transfer to a dust
grain \citep{DwekApJ1986}. If $E_f-E_i > E_\star$ it is not possible to 
jump from stage $i$ to stage $f$ through electron collisions; if
$E_f-E_i  < E_\star$ the jump is allowed for all electrons with energy
$E_f-E_i$ (which will transfer their entire energy to the grain) and for 
the electrons with energy $E_\mathrm{eff}$ (which will transfer an
energy $E_\mathrm{eff}\zeta(E_\mathrm{eff}) = E_f-E_i$; the function 
$\zeta(E)$ is the fraction of the electron energy which is deposited on 
a dust grain when $E> E_\star$, as defined in \citealt{DwekApJ1986}).

As in \citet{GuhathakurtaApJ1989} only cooling terms from level f+1 to 
level f are considered:
\[
A_{f,f+1}=\frac{4 \pi}{E_{f+1}-E_f}\int_0^\infty 
\pi a^2 Q_\mathrm{abs}(a,\lambda)
\;
B_\lambda(T_{\mathrm{d},f+1}) d\lambda,
\]
where $B_\lambda$ is the Planck function and $Q_\mathrm{abs}(a,\lambda)$ 
the absorption efficiency of a grain of radius $a$. Values for 
$Q_\mathrm{abs}(a,\lambda)$ were derived using the
\citeauthor{MieAnnPhys1908}'s (\citeyear{MieAnnPhys1908}) theory for
spherical dust grains, adopting the refractive 
index from the references in  Table~\ref{tab_stocha}.

\begin{figure}
\center{ \epsfig{file=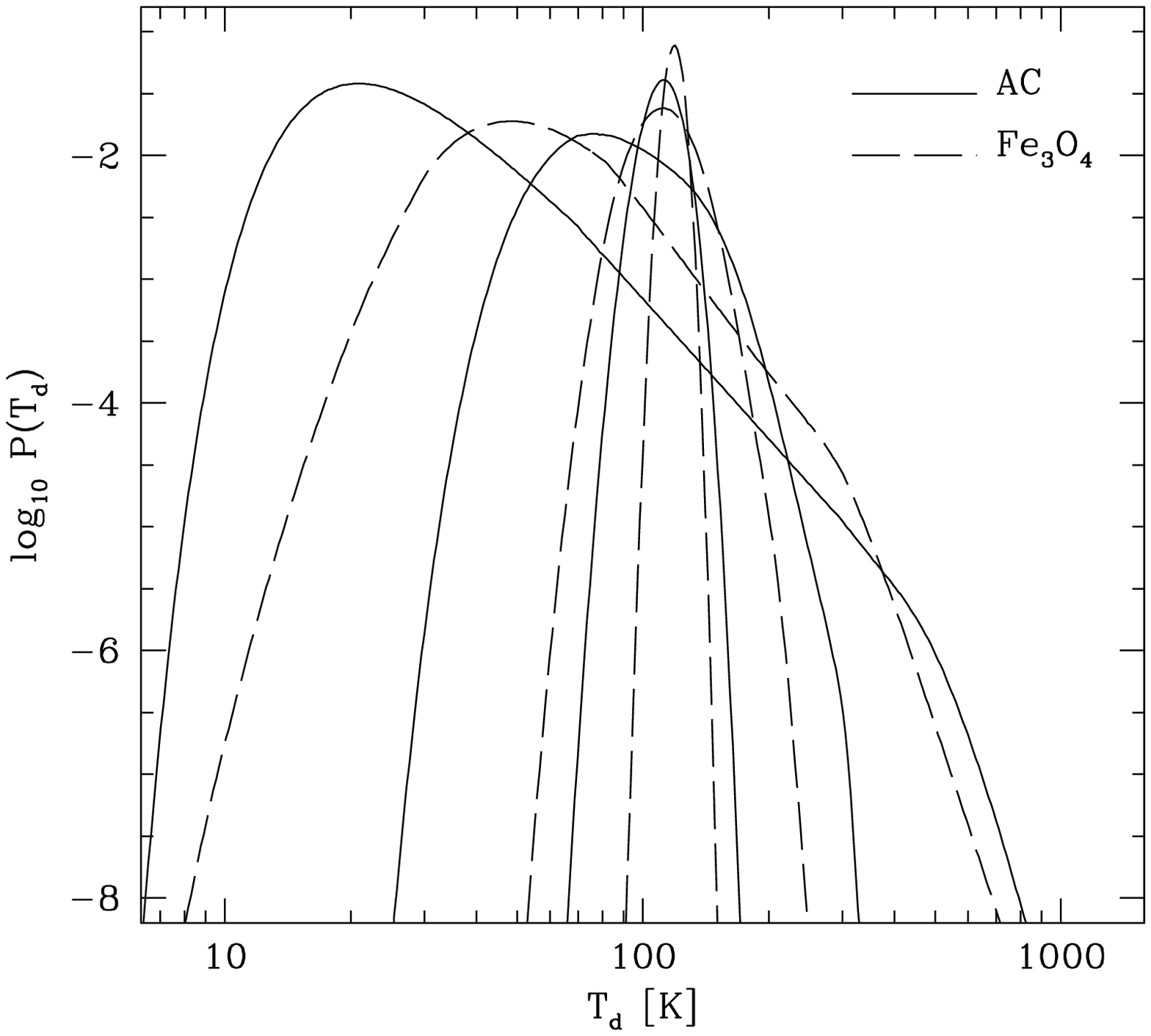,width=8cm} }
\caption{Temperature distributions $P(T_\mathrm{d})$ for AC and Fe$_3$O$_4$ 
dust grains in a hot gas with $T=10^8$K and $n=10$ electrons cm$^{-3}$. 
$P(T_\mathrm{d})$ is shown for radii $a=10$ \AA\ (broader distributions),
50 \AA\ and 200 \AA\ (distributions spanning a smaller range of $T_\mathrm{d}$).
}
\label{fig_stocha}
\end{figure}

In Fig.~\ref{fig_stocha} we show the temperature distributions for AC and 
Fe$_3$O$_4$ dust grains of radii $a$=10, 50 and 200\AA. Grains are exposed 
to a gas with $T=10^8$K and $n=10$ electrons cm$^{-3}$, typical conditions 
encountered in the ejecta swept by the reverse shock. Smaller grains have
broader temperature distributions, with a high temperature tail.

Using the sublimation rate of \citet{DraineApJ2002}, we have derived a
{\em sublimation temperature} $T_\mathrm{s}$, defined as the temperature 
necessary to completely consume a grain in about 20 yr, the typical time 
step of our simulation. For the grain sizes encountered in our work, we 
have $T_\mathrm{d}\ga$ 1200K. At any time in the ejecta evolution, 
only a negligible fraction of all grains would exceed that temperature.
As a result, dust sublimation is insignificant in our models.

\label{lastpage}

\end{document}